\documentstyle[12pt]{article}
\date{\today}

\begin{document}
\title{von Neumann Lattices in Finite Dimensions Hilbert spaces}
\author{M. Revzen*$^{\dag\ddag}$ and  F.$\;$ C.$\;$ Khanna*$^{\ddag}$\\
$\dag$  Department of Physics, Technion - Israel Institue of Technology,\\
Haifa 32000, Israel\\
$\ddag$  Theoretical Physics Institute,  Department of Physics, University of Alberta,\\
Edmonton,  Alberta, Canada  T6G 2J1.}

\maketitle

\begin{abstract}
The prime number decomposition of a finite dimensional Hilbert space reflects itself
in the representations that the space accommodates. The representations appear in
conjugate pairs for factorization to two relative prime factors which can be viewed
as two distinct degrees  freedom. These, Schwinger's quantum degrees of freedom, are
uniquely related to a von Neumann lattices in the phase space that characterizes the
Hilbert space and specifies the simultaneous definitions of
 both (modular) positions and (modular) momenta. The area in phase space for each
 quantum state in each of these quantum degrees of freedom, is shown to be exactly $h$,
 Planck's constant.

\end{abstract}

PACS: 03.67.Lx, 03.67. -a, 03.65.Ta

\section{Introduction}
Studies of finite dimensional Quantum Mechanics were undertaken in the early days of
the development of quantum mechanics by Weyl \cite{weyl} who showed thereby the
connection of the commutation relation to Schrodinger's wave equation. A systematic
study of this, finite dimensional quantum mechanics problem, was initiated 30 years
later  by Schwinger \cite{schwinger}. His work introduced to the field some of the
basic issues that are under intensive investigations in the most recent literature.
Among these are the relevance to the physics of finite dimensional quantum mechanics
of whether the dimensionality, M , is a prime number or is made up of product of
distinct primes. In the latter case, which is our main concern in the present work,
Schwinger \cite{schwinger} noted what he termed "quantum degrees of freedom" (see
also in \cite{bg}) which pertain to the  relatively prime number factors of M - these
are given a concrete representation in this paper.  He also considered sets of
complete and orthonormal bases that span the M- dimensional space and which are what
he termed of "maximum degree of incompatibility". Such vector bases under their
modern name of "mutually unbiased bases" (MUB)\cite{wootters,wootters1} or "conjugate
bases" \cite{wiesner, bennett} have led , with the recent developments in our
understanding of the foundation of quantum mechanics,  to intensive research in a
great variety of problems where implementations seems to involve finite dimensions.
These include the pioneering studies of von Neumann lattices and magnetic orbitals on
a finite phase space \cite{zak3},
 quantum measurements \cite{ekert},  teleportation \cite{paz} and others. A cogent review
 is given, e.g., by Vourdas \cite {vourdas}.\\

The kq representation was introduced by Zak \cite{zak1,zak2} in his study of the
continuum (infinite dimensional space) to conveniently handle electrons in a periodic
potential subjected to an external magnetic field. Zak's work was based on Weyl's
\cite {weyl} unitary displacement operators in both coordinates and momenta which,
for some particular choice of the parameters involve commuting operators whose
eigenvalues may be labelled by both the space position and the momentum of the
electron. (No violation of the uncertainty principle  incurs since the values of both
the coordinates and momenta are  modular,
 cf. \cite{aharonov}.) The Zak transform formalism proves of wide use also in signal
 processing problems \cite{janssen} where the corresponding modular variables are
 frequency and time.  The transform is of considerable theoretical interest in conjunction
 with it being parametrized by both the coordinate and momentum. Only recently \cite{berge}
 the conjugate basis to the kq rep was explicitly given. The finite dimensional kq
 representation was formulated in
 \cite{mann}. This can be done only for cases wherein the dimensionality, M, of the
 Hilbert space  is {\it not} prime \cite{mann,faqir,ady2}\footnote{strictly speaking
 factorizability of the dimensionality to  relatively prime factors is required for the
 definitions of the mapping from the discrete line to the discrete torus.}. For extended dimensionality
 (i.e. $M\; \rightarrow\;\infty$),
  already  at the very
 early stages of the development of quantum mechanics von Neumann \cite{vonneumann}
 suggested a physical way of accounting for states in phase space by discretizing the
 phase space with an area of $h$ (in the units $\hbar\;=\;1$ this equals $2\pi$) for
 each state. This problem was studied extensively, e.g. \cite{boon}.
 Recently it was shown \cite{ady} that on a finite phase plane the kq-coordinates and
 the sites on a von Neumann lattice are closely related.\\

In his studies of finite dimensional Quantum Mechanics, Schwinger \cite{schwinger,bg}
showed that if M, the dimensionality of the Hilbert space under consideration, is not
a prime or a power of a prime, then states in the space may be viewed as having
"quantum degrees of freedom". Thus, if $M\;=\;M_{1}M_{2}$ with $M_{1}\;,\;M_{2}$
relative primes, the state may be considered as accounting for two distinct degrees
of freedom one of dimension $M_{1}$ and the other of dimension $M_{2}$. This mapping
of the the one degree of freedom, M dimensional Hilbert space on a line (i.e. spanned
by $|q\rangle,;\;q\;=\;1,..,M$) onto two (Schwinger's "quantum") degrees of freedom
on a torus (spanned by $|q_1 q_2\rangle, \;\;q_1\;=\;1,..,M_1;\;q_2\;=\;1,..,M_2$)
and their conjugate representations (reps- henceforth) were introduced earlier
\cite{faqir}. These reps are reviewed herewith and the equivalence of these partially
localized state (PLS) and states over the von Neumann lattice is established. Thus we
show that a "quantum degree of freedom", much like a proper degree of freedom,
occupies a phase space area equals to $h$, Planck's constant. These states are
distinct form the finite dimension Zak's kq rep \cite{zak2} although when $M_1,\;M_2$
are co primes both bases are eigenstates of the commuting modular operators, one
pertaining to the position (q) and the other to the momentum (k). The latter (i.e.
Zak's) involve Bloch like symmetry in one of the variables. The phase relation
between Zak's reps (kq) and the reps used by us when both are applicable (i.e. when
the factors $M_1,\;M_2$ are co-primes) is given in the appendix. The present work
utilizes Schwinger's \cite{schwinger,bg} quantum degrees of freedom for finite
dimensional Hilbert space to consider states with two quantum degrees of freedom as
being labelled by the position of one and the momentum of the other.

The paper is organized as follows. The next two sections, section II and III outline
some previous results and briefly derive the formulae needed in our later discussion.
Section IV contains our main result, i.e.  relating partially localized states (PLS),
the kq reps states and states on von Neumann's lattices.  These relations bring to
the fore the role of the quantum degrees of freedom that were introduced by Schwinger
\cite{schwinger,bg} and give a new meaning to the localization in both the
coordinates and momenta in these variables (the price is complete delocalization in
their conjugate variables) and relate it to states over the von Neumann lattice. The
last section gives the conclusions and some remarks. The appendix gives the phase
relation among different representations.\\

\section{Mapping of Discrete Line to Discrete Torus}

 Schwinger \cite{schwinger}  showed that M-dimensional vector spaces  allow the
 construction of two unitary operators, U and V (in his
notation), which form a complete operator basis. This
means that if an operator commutes with both U and V it is, necessarily, a multiple
of the unit operator. These operators have a period
M, i.e.
\begin{equation} \label{period}
U^{M}\;=\;V^{M}\;=\;1,
\end{equation}
where M is the smallest integer for which this equality holds. The eigenvalues of both
U and V  are distinct: they are the M roots of
 unity,  i.e. with $|x\rangle$ the eigenfunction of U,
$$U|q\rangle\;=\;e^{i({2 \pi \over M})q}|q\rangle,\;\;\;|q+M\rangle\;=
\;|q\rangle,\;\;q=1,...,M.$$ The operator V is defined over these eigenvectors as
\begin{equation} \label{step}
V|q\rangle\;=\;|q-1\rangle.
\end{equation}
Schwinger then showed that the absolute value of the overlap between {\it any}
eigenfunction of U, $|q\rangle$  and any one of V, $|k\rangle,$  is a constant :
\begin{equation}
|\langle k|q\rangle|\;=\;{1 \over \sqrt M}.
\end{equation}
Vector bases with this attribute are referred \cite{wootters} to as mutually unbiased
 (MUB) or conjugate vector bases (we use these terms interchangeably).\\

A specific example of the M-dimensional space is the following: consider  M points
 on a line, i.e.,
consider discretized and truncated spatial coordinate x and its conjugate momentum p
as our M-dimensional space. This  may be realized by imposing boundary conditions on
the spatial coordinate, x, of the wave functions under study, $\psi(x),$ and on their
Fourier transforms, F(p) (we take $\hbar=1$) \cite{zak2}:
$$\psi(x\;+\;Mc)\;=\;\psi(x),\;\;F(p\;+\;{2\pi \over c})\;=\;F(p).$$
Here M is an integer - it is the dimensionality of the Hilbert space, and we term c
the ``quantization length''.  As a consequence of the above boundary conditions we
have that the value of the spatial coordinate, x, and the value of the momentum, p,
are discrete and finite:
$$x\;=\;sc,\;s\;=\;1,...,M;\;\;p\;=\;{2\pi \over Mc}t,\;t\;=\;1,...,M.$$
In this case we may replace the operators x and p by the unitary operators
\begin{equation} \label{Tt}
\tau(M)\;=\;e^{i({2\pi \over Mc})x};\;T(c)\;=\;e^{ipc}.
\end{equation}
These operators satisfy the basic commutator relation
\begin{equation}
\tau(M)T(c)\;=\;T(c)\tau(M)e^{-i{2\pi \over M}}.
\end{equation}
They exhibit the dimensionality (i.e. periodicity) automatically
\begin{equation}
[\tau(M)]^{M}\;=\;[T(c)]^{M}\;=\;1,
\end{equation}
and we may associate Schwinger's operator $U$ with $\tau(M)$ and his  $V$ with $T(c)$
(henceforth we take c=1).\\

For our analysis where we take M as as factorizable: $M\;=\;M_{1} \cdotp M_{2},$ it
is convenient to represent the number M in terms of prime numbers, $P_{j}$,
\begin{equation}
M\;=\;\prod_{j=1}^{N}P_{j}^{n_{j}},\;\;P_{j}\;\neq \;P_{i},\;j\;\neq \;i,
\end{equation}
where the $n_{j}$ are integers, and more concisely we denote
$P_{j}^{n_{j}}\;{\rm by}\;m_{j},$ i.e.
\begin{equation} \label{primes}
M\;=\;\prod_{j=1}^{N}m_{j}.
\end{equation}
We find thus that the greatest common divisor (gcd) among the
$m_{j}s$ is 1:
\begin{equation} \label{gcd}
gcd(m_{j},m_{i})\;=\;1,\;\forall\;j\;\neq \;i,
\end{equation}
i.e. distinct $m_{i}s$ are relatively prime. In our study we consider bi partitioning
of the product that represents M (Eq. (\ref{primes})) into two factors,
\begin{equation}\label{product}
M\;=\;M_{1}M_{2}.
\end{equation}
Here $M_{1}$ incorporates one part of the N factors of Eq. (\ref{primes}) and $M_{2}$
contains the other part. Our way of bi-partitioning implies that the two numbers,
$M_{1}\;{\rm and}\; M_{2},$ are relatively prime, viz. $gcd(M_{1},M_{2})\;=\;1$. In
our discussion of the kq representation \cite{zak1,mann,faqir} the above was used to
show that the number of kq representations, $\chi(M),$ each with its conjugate, which
form a complete basis can be accommodated in the M dimensional space, is simply
related to the number of primes, N,  that appear in M :
\begin{equation}\label{chi}
\chi(M)\;=\;2^{N-1}.
\end{equation}
(It should be noted that the familiar finite dimensional Fourier representation is
included in this counting.) The mapping of the M dimensional, one degree of freedom
(a line) to (fake) two degrees of freedom (a torus) whose dimensions are $M_{1}$ and
$M_{2}$ can be accomplished in $\chi(M)\;-1$ ways (the finite dimensional Fourier
transformation should not be included). We now introduce:

\begin{equation}\label{L}
L_{1}\;=\;{M\over M_{1}}\;=\;M_2\;;\;\;L_{2}\;=\;{M \over M_{2}}\;=\;M_1.
\end{equation}

This implies that the equation ,
\begin{equation} \label{ts}
q\;=\;q_{1}L_{1}\;+\;q_{2}L_{2}\;\;[mod\;M];\;\;q=1,..,M;\;q_{1}=1,...,M_{1};\;
q_{2}=1,...,M_{2},
\end{equation}
has a unique solution $q$ for every pair $[q_{1},q_{2}],$ with q running over its
whole range of M values.  We will now \cite {schwinger,ekert,vourdas} modify Eq.
(\ref{ts}) to attain this simpler relation among the solutions. This is obtained by
applying the Chinese remainder theorem \cite{ekert} to the solution of the two
congruences
\begin{eqnarray}
q\;&=&\;q_{1}\;\;[mod\;M_{1}] \nonumber\\
q\;&=&\;q_{2}\;\;[mod\;M_{2}].
\end{eqnarray}
The solution of these is
\begin{equation}
q\;=\;q_{1}N_{1}L_{1}\;+\;q_{2}N_{2}L_{2} \;\;[mod\;M],
\end{equation}
with
\begin{equation}\label{inverses}
N_{2}\;=\;L_{2}^{-1}\;[mod\;M_{2}],\;{\rm and}\;N_{1}\;=\;L_{1}^{-1}\;[mod\;M_{1}].
\end{equation}
For example, with M = 15, we have
\begin{equation}\label{example}
L_{1}\;=\;5;\;\;L_{2}\;=\;3\;\rightarrow\;N_{1}\;=\;N_{2}\;=\;2.
\end{equation}
We have then,
\begin{equation}
e^{i{2 \pi \over M}}\;=\;e^{i{2 \pi \over M_{1}}N_{1}}e^{i{2 \pi \over M_{2}}N_{2}}\;
\rightarrow\;e^{i{2 \pi \over M}x}\;=\;e^{i{2 \pi \over M_{1}}N_{1}x}e^{i{2 \pi \over
M_{2}}N_{2}x},
\end{equation}
and,
$$e^{ip}\;=\;e^{ipN_{1}L_{1}} e^{ipN_{2}L_{2}}.$$
Further, we may label

$$|q\rangle\;=\;|q_{1}N_{1}L_{1}\;+\;q_{2}N_{2}L_{2}\rangle;$$
$$|k\rangle\;=\;|k_{1}N_{1}L_{1}\;+\;k_{2}N_{2}L_{2}\rangle,$$
We have thus that the (complete) operator basis pair may be replaced by the two pairs
(note the removal of $N_{i}$ from the V terms) as follows:
\begin{eqnarray}\label{mapp}
\tau(M)= e^{i{2 \pi \over M} x}\;&\rightarrow&\;
 \tau(M_{1})\tau(M_{2})= e^{i{2\pi \over M_{1}}x}e^{i{2 \pi \over M_{2}}x}\nonumber
 \\
 T(1)= e^{ip};&\rightarrow&\;,\;T(L_{1})T(L_{2})=e^{ipL_{1}}e^{ipL_{2}}.
\end{eqnarray}

 While the basis
$|q\rangle$ may be expressed via $|q_{1}\rangle|q_{2}\rangle$ with
\begin{equation}
e^{i{2 \pi \over M_{i}}x}|q_{i}\rangle\;=\; e^{i{2 \pi \over
M_{i}}q_{i}}|q_{i}\rangle;\;\;e^{ipL_{i}}|k_{i}\rangle\;=\;e^{i{2 \pi \over
M_{i}}k_{i}}|k_{i}\rangle.
\end{equation}
Defining
\begin{eqnarray}
\Delta^{M_{i}}(x)\;&=&\;1,\;x\;=\;0\;mode[M_{i}]\nonumber \\
                   &=&\;0\;\;otherwise, \nonumber
\end{eqnarray}
allows equating,
\begin{eqnarray}
<q|q_{1}N_{1}L_{1}\;+\;q_{2}N_{2}L_{2}>\;&=&\;\Delta^{M}(q\;-\;q_{1}N_{1}L_{1}\;-\;q_{2}N_{2}L_{2})\;=
\nonumber \\
\Delta^{M_{1}}(q\;-\;q_{1})\Delta^{M_{2}}(q\;-\;q_{2})\;&=&\;<q|[|q_{1}\rangle|q_{2}\rangle].
\end{eqnarray}
And,
\begin{eqnarray}
<k|k_{1}N_{1}L_{1}\;+\;k_{2}N_{2}L_{2}>\;&=&\;\Delta^{M}(k\;-\;k_{1}N_{1}L_{1}\;-\;k_{2}N_{2}L_{2})\;=
\nonumber \\
\Delta^{M_{1}}(k\;-\;k_{1})\Delta^{M_{2}}(k\;-\;k_{2})\;&=&\;<k|[|k_{1}\rangle|k_{2}\rangle].
\end{eqnarray}

The formulae
\begin{eqnarray}\label{qq}
|q\rangle\;&\rightarrow&\;|q_{1}\rangle |q_{2}\rangle \nonumber \\
|k\rangle\;&\rightarrow&\;|k_{1}\rangle ||k_{2}\rangle
\end{eqnarray}
with \ref{mapp} constitutes the transcription of the M dimensional, one degree of
freedom system to the (Schwinger's quantum) two degrees of freedom one of $M_{1}$ and
the other of $M_{2}$ dimensions ($M\;=\;M_{1}M_{2})$ which is our mapping of the M
dimensional line to a torus. Our numerical subscripts designates the dimensionality;
thus, e.g., $|q_1\rangle$ refers to the spatial coordinate with $q_1\;=\;1,2,...M_1.$  \\
We shall now discuss briefly some attributes of what is termed \cite{faqir} the
$|q_1\rangle|q_2\rangle$ representation (rep henceforth). The conjugate basis viz
$|k_1 \rangle|k_2 \rangle$ may be evaluated directly via the result
$$\langle q|k \rangle\;=\;{e^{iqk} \over \sqrt{M}}$$ to give
\begin{equation}\label{q-q}
\langle q_1|\langle q_2||k_1 \rangle|k_2
\rangle\;=\;{e^{i(q_{1}k_{1}L_{1}+q_{2}k_{2}L_{2}){2 \pi \over M}} \over \sqrt M},
\end{equation}
and, by direct evaluation
\begin{equation}
\langle q_i|k_i \rangle\;=\;{e^{i{2\pi \over M_i}N_i q_i k_i} \over \sqrt
M_i}\;\;i=1,2.
\end{equation}

   Within the
  description in terms of two (quantum) degrees of freedom (QDF). E.g. when accounting for
$|q\rangle$ via $|q_1 \rangle|q_2 \rangle$ (cf. Eq.(\ref{qq})) we may construct
  eigen functions the above operators by Fourier transformation in either one of the
  variables (we show in the appendix that we need not differentiate between the two
  forms)
  \begin{eqnarray}
  |q_1,k_2;\rangle\;=\;|q_1\rangle |k_2 \rangle\;&=&\;{1 \over \sqrt
  {M_2}}\Sigma_{q_2}e^{i{2 \pi \over M_{2}}k_{2}q_{2}N_{2}}|q_{1}\rangle |q_2 \rangle,
  \nonumber \\
                                                  &=&{1 \over \sqrt M_{1}}
\Sigma_{k_1}e^{-i{2 \pi \over M_{1}}k_{1}q_{1}N_{1}}|k_{1}\rangle|k_{2}\rangle.
\end{eqnarray}
One can readily check that these are indeed the eigenfuctions of the operators
Eq.(\ref{ops}) with the same eigenvalues. In this form the state $|q_1,k_2;\rangle$
can be described as partially localized (PLS) as we have
\begin{equation}\label{pls}
\langle q_{1}|\langle
q_{2}|q_{01},k_{02};\rangle\;=\;{\Delta^{M_1}(q_1\;-\;q_{01})e^{i{2 \pi \over
M_{2}}k_{2}q_{2}N_{2}} \over \sqrt M_2},
\end{equation}
thus it is localized in the $q_1$ variable while completely delocalized in the $q_2$
variable. (In \cite{mann} we used $e^{ipM_{2}}$ instead of the present
$T(N_{1}L_{1});$ these
 two operators have the same eigenvalues and eigenstates,
but enumerated differently.)\\
In the next section we expand our presentation, \cite{ady}, to show that these states
are states over the von Neumann lattice.

\section{von Neumann lattices}

We now consider the von Neumann lattice for the M dimensional Hilbert space. To this
end we first describe the phase space of our system: The spatial like coordinates,
 q, are now discrete, $q\;=\;0,1,....,M-1$ ( this in dimensionless units: $q\;=\;cM$
 is the "size" of
  our Hilbert space)  and label the eigenfunction $|q\rangle$  of the operator
  $exp[i({2\pi \over M})x]$. These label the abscissa of our phase space.
  The momentum like coordinates are $p\;=\;{2\pi \over M}k;\;\;k\;=\;0,1,....,M-1,$
 the label k of $|k \rangle$, the eigenfunctions of $exp[ip]$  are used for
the ordinate of our phase space.\\

We now consider the case wherein $M\;=\;M_{1}\cdot M_{2},\;\;gcd[M_{1},M_{2}]\;=\;1.$
Since as k runs from 1 to M the momenta runs from
  $p\;=\;{2 \pi \over Mc}\hbar\; to\; {2 \pi \over c}\hbar,$  the "distance" $\delta k\;=\;1$
  represents $\delta p\;=\;{2 \pi \over Mc}\hbar.$ (In dimensionless units -
  i.e. $c\;=\;\hbar\;=1$ - p runs from ${2 \pi \over M}\;to\;{2 \pi }\; and\;
   \delta p\;=\;{2 \pi \over M}.$) The "distance" $\delta q\;=\;1$ along the abscissa
   is, obviously, c ( 1 in dimensionless units). In this way we see that each "point"
   marked on our phase space by (q,k) may be viewed as representing an "area",
$\delta q \delta k \;=\; {2 \pi \over M}.$\\
Our aim in this section is to show that $|q_1,k_2\rangle$  states, i.e.  the eigen
functions of
$$e^{i[{2\pi \over M_{1}}x]}e^{i[pM_{1}]}$$ as given above, Eq. (\ref{ops}) have the
attractive physical properties of being states over the von Neumann lattice, a term
that will also defined herewith.\\
  A von Neumann lattice in this phase space are the M points whose coordinates are
  \begin{equation}
  q\;=\;nM_{1},\;k\;=\;mM_{2},\;n\;=\;0,1,2...,M_2 -1,\;m\;=\;0,1,2...M_1 -1.
  \end{equation}
A {\it displaced} von Neumann lattice are the M points - here the points are defined
mode[M].
\begin{eqnarray}
  q\;&=&\;q_{01}\;+\;nM_{1};\;\;n,k_2\;=\;0,1,2...,M_2 -1,\nonumber \\
  k\;&=&\;k_{02}\;+\;\;mM_{2};\;\;m,q_{01}\;=\;0,1,2...M_1 -1.
  \end{eqnarray}
We now define a "state over a von Neumann lattice" to be the state whose density
matrix representative, $\rho$ is given by
\begin{eqnarray}
|\langle q|\rho|k\rangle|\;&=&\;{1 \over \sqrt{M}}\;\;on\;a\;lattice\;point \nonumber\\
                           &=&\;0\;\;otherwise.
\end{eqnarray}
(This requirement implies that only M out of the $M^2$ matrix elements are non
vanishing.) It is easily verified that the partially localized state (PLS)  Eq.
(\ref{pls}) with localization coordinates at the origin ($q_{01}\;=\;k_{02}\;=\;0$)
is such a state (summation over repeated indices is implied):
\begin{eqnarray}
\langle q|\rho |k \rangle \;&=&\;\langle q|q_1 q_2 \rangle \langle q_1 q_2|\psi
\rangle
\langle \psi|k_1 k_2 \rangle \langle k_1 k_2 |p \rangle  \nonumber \\
                            &=&\Delta(x-q_{1}N_{1}L_{1}-q_{2}N_{2}L_{2}){\Delta^{M_1}(q_{1})
                            \over \sqrt {M_{2}}}{\Delta^{M_2}(k_{2}) \over
                            \sqrt {M_{1}}}\Delta (p-k_{1}L_{1}-k_{2}L_{2}).
\end{eqnarray}
Thus the PLS is a state over a von Neumann lattice. For example, M = 15,
$M_{1}\;=\;3,\;M_{2}\;=\;5; \;L_{2}\;=\;3,\;N_{2}\;=\;2$ leads to non vanishing matrix
elements for
$$q\;=\;q_{2}N_{2}L_{2}\;\rightarrow\;q\;=\;0,6,12,3,9;\;\;k\;=\;k_{1}L_{1}\;\rightarrow
\;k\;=\;0,5,10.$$ We take it as obvious that with each von Neumann lattice we have
its conjugate obtained
 by interchanging q with k. \\
To study the case of displaced von Neumann lattices we  consider briefly the PLS with
(at least) one coordinate not at the origin. For example
\begin{equation}
\langle q_1 q_2 |\psi \rangle\;=\; {\Delta^{M_1}(q_{1}-q_{10})e^{i[q_{2}k_{20} {2\pi
\over M_{2}}]} \over \sqrt M_{2}};\;\;\;q_{01},\;k_{02}\;\ne\;0.
\end{equation}
The non vanishing elements of the density matrix are now the M points
\begin{eqnarray}
q\;&=&\;q_{01}N_{1}L_{1}\;+\;q_{2}N_{2}L_{2};\;\;q_{2}\;=\;0,...,M_{2}-1; \nonumber \\
k\;&=&\;k_{1}L_{1}\;+\;k_{02}L_{2},\;\;k_{1};\;=\;0,...M_{1}-1.
\end{eqnarray}
This is a von Neumann lattice with its origin shifted to $
q_{01}N_{1}L_{1}\;,k_{02}L_{2}$. Note that corresponding to PLS
$|q_{01},k_{02}\rangle$ the non vanishing terms involve the coordinates of the
conjugate state, $|k_1,q_2\rangle$. This corresponds to a localized state at x say
implies all momentum state equally probable.
 This state is orthogonal to the von Neumann state considered above. Indeed each of the
 M pairs of coordinates $q_{01},k_{02}$ represents a state over the von Neumann lattice
 with its origin shifted to the designated phase space point. These states are easily
 shown to be  orthogonal. Thus, considering the overlap between  arbitrary  such states
(summation convention implied)
\begin{eqnarray}
\langle \psi_{1}|\psi_{2}\rangle\;&=&\;\langle \psi_{1}|q_{1}q_{2}\rangle
\langle q_{1}q_{2}|\psi_{2} \rangle \nonumber \\
                                  &=&\;{\Delta^{M_1}(q_{1}-q_{01}) e^{i[q_{2}k_{02}
                                  {2\pi \over M_{2}}]} \over \sqrt {M_{2}}}
                                  {\Delta^{M_1}(q_{1}-q'_{01}) e^{-i[q_{2}k'_{02}
                                  {2\pi \over M_{2}}]} \over \sqrt {M_{2}}} \nonumber \\
                                  &=&\;\Delta^{M_1}(q_{01}-q'_{01}) \Delta^{M_2}(k_{02}-k'_{02}).
\end{eqnarray}
Thus these M states span the space and form a complete orthonormal basis. We shall
show below that these are the PLS considered above. In closing this section we wish
to emphasize that by construction each von Neumann lattice point represents an area
of $2\pi$ in our phase space diagram. The total number of such points within a
rectangle labelled by $(q_{01},k_{02})$ is M. Thus, e.g. for the rectangle labeled by
  $(q_{01}=0,k_{02}=0)$  includes the  phase space coordinates within the  area
  contained in the rectangle  $(0,0);\;(M_{1},0);\;(0,M_{2});\;(M_{1},M_{2}).$ There
  are M possible values for $\langle x|\rho | p \rangle$ within this rectangle.
  Thus the $2\pi$ area that contain M points - each labels a state over the von Neumann
  lattice (there are M such states) each state includes M "points": one in  each  rectangle
  - this point is discussed further below.\\

We now show that PLS are kq states, i.e. are the eigen functions of
$$e^{i[{2\pi \over M_{1}}x]}e^{i[pM_{1}]}$$ (repeated indices are summed over):
\begin{eqnarray}
\langle q_{1}q_{2}|\;e^{i[{2\pi \over M_{1}}x]}e^{i[pM_{1}]}|\psi \rangle\;&=&\;
                  e^{i[{2\pi \over M_{1}}q_{1}]}\langle q_{1}q_{2}|k_{1}k_{2}\rangle
                   e^{i[{2\pi \over M_{2}}k_{2}]}\langle k_{1}k_{2}|\psi \rangle \nonumber \\
                  &=&\;e^{i[{2\pi \over M_{1}}q_{1}]}e^{i[{2\pi \over M_{2}}k_{2}]}
                  \langle q_{1}q_{2}|\psi \rangle.
\end{eqnarray}

We have then that the $|k_{1}q_{2}\rangle$ rep (as well as their concomitant
conjugates, $|k_{2}q_{1}\rangle$), PLS (at the corresponding points) and the states
over the von Neumann lattice (shifted to the corresponding spots), are one and the
same states. (We note that these states differ from Zak's \cite{zak1} states.  The
latter are defined regardless whether $M_{1},\;M_{2}$ are co primes or not.) The
completeness and orthogonality of the $|k_{1}q_{2}\rangle$ states assures their
validity for the other two. The states over the von Neumann lattice were shown to
occupy precisely an area of $h$, Planck's constant in phase space for the finite
dimensional case. While the coordinates of the non vanishing density matrix for the
states (partially) localized at $k_{02}\;=\;q_{01}\;=\;0$ were at the appropriate von
Neumann lattice (as well as their conjugates $k_{01}\;=\;q_{02}\;=\;0$ ). We thereby
accounted for the connection between the $q_1,q_2$ reps
and von Neumann lattices that were obtained recently \cite{ady}.\\

We now return to the more general case: we consider an arbitrary pair with partially
 localized phase points: $q_{01},\;k_{02}$. These, in turn, lead to different marked
points  in our phase space - it specifies a  lattice point within the rectangles
considered above and may serve as a label for a partially localized state localized
at $q_{01}$ and having quasi momenta $k_{02}.$ One can readily verify that this state
may be viewed as a state over the von Neumann lattice where the M points for which
the the matrix elements $$\langle k|q_{01},\;k_{02}\rangle \langle q_{01},\;k_{02}|q
\rangle$$ do not vanish are given by,
\begin{equation}
q\;=\;q_{01}\;+\;nM_{1},\;k\;=\;k_{02}\;+\;mM_{2},\;n\;=\;0,1,2...,M_2 -1,\;
m\;=\;0,1,2...M_1 -1.
\end{equation}
 Thus these states are states over the von Neumann lattice
with the whole lattice origin being shifted to the point $q_{01},k_{02}.$\\
In closing this section we would like to point out that the results above are based
on the particular choice of labelling, viz. on having achieved a mapping of the M
dimensional line geometry ($q\;=\;1,2....,M)$ to a torus like geometry with one torus
radius $M_{1}$ dimensional ($q_{1}\;=\;1,2,...M_{1})$ and the other $M_{2}$
dimensional ($q_{2}\;=\;1,2,...M_{2})$. This was shown to be possible for
$M\;=\;M_{1}M_{2}$ with $gcd [M_{1},M_{2}]\;=\;1.$

\
\section{Concluding Remarks}

In a finite, M dimensional Hilbert space with $M\;=\;M_1 M_2$ and $M_1,\;M_2$ are co
primes i.e. $g.c.d\;[M_1,M_2]\;=\;1,$ one may, ala Schwinger, view the states as made
of two quantum degrees of freedom (QDF):
 one of dimensionality $M_1$ and the other $M_2$. This was utilized to consider the
 $|q_1,q_2\rangle$ with $q_1$ being $M_1$, while $q_2$ the  $M_2$ dimensional coordinates.
 We analyzed the two QDF states as eigenfunctions of the two commuting operators
 $e^{ipM_2},\;e^{i{2\pi \over {M_2}}x},$ which, in the present context may be viewed
 as referring to the two QDF. We showed that the the eigenstates, of these operators
 may be obtained via Fourier transformation in one of the QDF.    e.g.
 $|k_{1},q_2;C\rangle$ is the
 Fourier like transform in one of these QDF variables
($q_1$ in this example). (Either of these sets spanning the M dimensional space under
study.)

 Thus the basis $|k_1,q_2;C \rangle$  specify $k_1$ as the (modular)
 momenta of
the $M_1$ dimensional QDF and $q_2$ the (modular) position of the second,
$[mode\;M_2],$ variable.(Hence the conjugate basis is $|k_2,q_1 \rangle$ with
analogous meaning for the subscripted labels.) The M orthogonal basis vectors span
the space - as does the conjugate basis. These states are termed partially localized
states as therein (in the above example) the second QDF is localized while the first
has its momenta defined. We termed these states partially localized states (PLS). We
discuss briefly the distinction of these from Zak's kq states in the appendix.
In this, finite dimensional quantum mechanics, one may draw the phase space of the
system as a whole (i.e. M dimensional) by marking the abscissa by the eigenvalues of
$\exp[i{2\pi \over M}x]$ by the discrete, position like, eigenvalue of the state
$|q\rangle$, ($q=0,1,2,...,M-1$), and the ordinate by eigenvalues of $\exp[ip]$ by
the momentum like eigenvalue of the state, $|p\rangle$ , ($p={2\pi \over
M}k;\;k=0,1,2,...,M-1 .$) (Giving thereby $M^2$ points forming a square, with each
point designating an area of $2\pi \over M.$) The von Neumann lattice in this phase
space and pertaining to the factorization $M\;=\;M_1 M_2;\;g.c.d\;[M_1,M_2]\;=\;1,$
is given by the points
$$q\;=\;nM_2,\;\;k\;=\;mM_1;\;n=0,1,..M_1 -1,\;\;m=0,1,..M_2 -1.$$ We refer to these
points as vn points. Clearly each vn point designates an area containing M phase space
points i.e. it designates an area of $2\pi.$ (Recall that we work in units of $\hbar\;=\;1$,
 i.e. the area in more physical units is h, Planck's constant.) We then defined "states
 over the von Neumann lattice" as those density matrices whose only non vanishing matrix
 elements are
$$\langle x|\rho|k \rangle \;=\;{1 \over \sqrt M};\;\;x,\;p\;{\rm{on\;von\;Neumann\;lattice}}.
$$
Since the total number of von Neumann lattice points is M and each point occupies and
area of ${2\pi \over M}$ each von Neumann state occupies exactly an area of $2\pi$
(i.e. an area of h, Planck's constant). We then showed that the state $|k_1 =0,q_2
=0\rangle$ (as well as its conjugate state) are states over von Neumann lattice and
thus occupy an area of h, Planck's constant. This led to the demonstration  that
concomitant to this state over the von Neumann lattice each of the other vectors in
this basis designates a shifted von Neumann lattice - the shift being by the
coordinates of the designated vector. Thus the vector $k_1 = k_{01}\ne 0,\;q_2 =
q_{02}\ne 0$ is a state over the von Neumann lattice shifted to
$(k_{01}\;[mode\;M_{1}],q_{02}\;[mode\;M_{2}]).$ Here too the area occupied by each
of the M points is ${2\pi \over M}.$\\ Thus the states over von Neumann lattice are
labelled by the the coordinates of one of the QDF and by the momentum of the other.

\bigskip
{\bf Acknowledgements}\\
FCK acknowleges the support of NSERC. MR acknowledes numerous informative discussions
with Professors Joshua Zak and Ady Mann and The Theoretical Physics Institute for
partial support.
\bigskip
$ \ast $ {\bf electronic addresses:} revzen@physics.technion.ac.il,
khanna@phys.ualberta.ca,\\
ady@physics.technion.ac.il,  zak@physics.technion.ac.il.

\title{\bf  Appendix: Phase relation among different  reps}\\

\maketitle

Utilizing the general Definitions, Eq.(\ref{product}) to Eq.\ref{q-q}) one obtains
directly





\begin{eqnarray}
<q|q_{1}N_{1}L_{1}\;+\;q_{2}N_{2}L_{2}>\;&=&\;\Delta^{M}(q\;-\;q_{1}N_{1}L_{1}\;-\;q_{2}N_{2}L_{2})\;=
\nonumber \\
\Delta^{M_{1}}(q\;-\;q_{1})\Delta^{M_{2}}(q\;-\;q_{2})\;&=&\;<q|[|q_{1}>|q_{2}>],\nonumber
\\
<k|k_{1}N_{1}L_{1}\;+\;k_{2}N_{2}L_{2}>\;&=&\;\Delta^{M}(k\;-\;k_{1}N_{1}L_{1}\;-\;k_{2}N_{2}L_{2})\;=
\nonumber \\
\Delta^{M_{1}}(k\;-\;k_{1})\Delta^{M_{2}}(k\;-\;k_{2})\;&=&\;<k|[|k_{1}>|k_{2}>].
\end{eqnarray}
We also have
\begin{eqnarray}
<q'_1\;+\;q'_{2}L_2|q_1N_1L_1\;+q_2N_2L_2>\;&=&\;\Delta(q'_1\;-\;q_1N_1L_1\;-\;(q_2-q'_2N^{-1}_2)N_2L_2)
\nonumber \\
                                            &=&\;\Delta^{M_1}(q'_1-q_1)\Delta^{M_2}(q'_1-q_2+q'_2N^{-1}_2).
                                            \nonumber
\end{eqnarray}
 The above implies also shift operators as follows:
 $$e^{i{2 \pi \over M}x}|k>\;=\;|k+1>;\;\;e^{ip}|q>\;=\;|q-1>.$$
 $$e^{i{2 \pi \over
 M_{1}}x}|k_{1}>\;=\;|k_{1}\;+\;M_{2}>;\;\;e^{ipL_{1}}|q_{1}>\;=\;|q_1\;-\;M_{2}>.$$
 With similar equations for $|k_2>,\;|q_2>.$\\

We now consider the wave functions explicitly \\

 Define\\
\begin{eqnarray}\label{c1}
|q_1 k_2;C_{1}>\;&\equiv& \;{1 \over \sqrt M_{1}} \Sigma_{k_1}e^{-i{2 \pi \over
M_{1}} k_{1}q_{1}N_{1}}|k_{1}N_{1}L_{1}\;+\;k_{2}N_{2}L_{2}> \nonumber \\
                 &=&\;{1 \over\sqrt M_{1}} \Sigma _{k_1}e^{-i{2 \pi \over
M_{1}}k_{1}q_{1}N_{1}}|k_{1}>|k_{2}>=|q_1>|k_2>.
\end{eqnarray}
We have by direct calculations that (the eigenvalue indices are deleted for brevity),
$$e^{i{2 \pi \over M_{1}}x}|C_{1}>\;=\;e^{i{2 \pi \over
M_{1}}q_{1}}|C_{1}>;\;\;e^{ipL_{2}}|C_{1}>\;=\;e^{i{2 \pi \over
M_{2}}k_{2}}|C_{1}>.$$

And, in similar fashion, define
\begin{equation}
|q_1 k_2;C_{2}>\;\equiv \;{1 \over \sqrt M_{2}}\Sigma_{q_{2}}e^{i{2 \pi \over
M_{2}}k_{2}q_{2}N_{2}}|q_{1}N_{1}L_{1}\;+\;q_{2}N_{2}L_{2}>.
\end{equation}
With,
$$e^{i{2 \pi \over M_{1}}x}|q_1 k_2;C_{2}>\;=\;e^{i{2 \pi \over
M_{1}}q_{1}}|q_1 k_2;C_{2}>;\;\;e^{ipL_{2}}|C_{2}>\;=\;e^{i{2 \pi \over
M_{2}}k_{2}}|q_1 k_2;C_{2}>.$$

Thus $|q_1 k_2;C_{1}>$ and $|q_1 k_2;C_{2}>$ are eigenfunctions of the complete set
of commuting operators for the M dimensional space under study. To calculate the
(possible) phase difference between them we evaluate the overlap,
\begin{equation}\label{phased}
<q_1 k_2;C'_{1}|q_1
k_2;C_{2}>\;=\;\Delta^{M_{1}}(q'_{1}\;-\;q_{1})\Delta^{M_{2}}(k'_{2}\;-\;k_{2}),
\end{equation}
This evaluation requires  the evaluation of the Fourier transform, $\langle k_1|q_1 \rangle$:\\

\begin{eqnarray}
<k_{1}|e^{i{2 \pi \over M_{1}}x}|q_{1}>\;&=&\;e^{i{2 \pi \over
M_{1}}q_{1}}<k_{1}|q_{1}>\;=\;<k_{1}\;-\;M_{2}|q_{1}>, \nonumber \\
<k_{1}|e^{ipL_{1}}|q_{1}>\;&=&\;e^{i{2 \pi \over
M_{1}}k_{1}}<k_{1}|q_{1}>\;=\;<k_{1}|q_{1}\;-\;M_{2}>.\nonumber
\end{eqnarray}
This leads to (with the proper normalization, see Eq.(\ref {c1})).
$${1 \over \sqrt M_{1}}e^{-i{2 \pi \over M_{1}}q_{1}k_{1}N_{1}}\;=\;<k_{1}|q_{1}>.$$
We check this by evaluating (we do not include the normalization for simplicity),
\begin{eqnarray}
<k|q>\;&=&\;e^{-i{2 \pi \over M}qk} \nonumber \\
       &=&\;e^{-i{2 \pi \over
       M}(q_{1}N_{1}L_{1}\;+\;q_{2}N_{2}L_{2})(k_{1}N_{1}L_{1}\;+\;k_{2}N_{2}L_{2})}\nonumber
       \\
       &=&\;e^{-i{2 \pi \over M_{1}}q_{1}k_{1}N_{1}}e^{-i{2 \pi \over
       M_{2}}q_{2}k_{2}N_{2}}\nonumber \\
       &=&\;<k_{1}|q_{1}><k_{2}|q_{2}>.
\end{eqnarray}

The kq rep for finite dimensional were considered by Zak \cite{zak1,mann} and are
defined as the eigenfunctions of the (commuting) operators
\begin{equation} \label{ops}
\tau(M_{1})\;=\;e^{i({2 \pi \over M_{1}})x};\;T(L_{2})\;=\;e^{ipL_{2}},
\end{equation}
which  are periodic in one variables and quasi periodic in the other. These can be
written in terms of $|q\rangle,$ the eigen functions of $\tau(M)\;=\;e^{i({2 \pi
\over M})x}$ (we label these by $E_{1}$) as
\begin{equation}
|q_1,k_2;E_1\rangle\;\equiv \; {1 \over \sqrt M_{2}}\sum_{q_{2}}e^{i{2 \pi \over
M_{2}}k_{2}q_{2}}|q_{1}\;+\;q_{2}L_{2}\rangle.
\end{equation}

In terms of the eigen functions of $T(L_2)$ these can be expressed via (the extra
label here is $E_2$),

\begin{equation}
|q_1,k_2;E_2>\;\equiv \; { 1 \over \sqrt M_{1}}\Sigma_{k_1}\;e^{-i{2 \pi \over
M_{1}}k_{1}q_{1}}|k_{2}\;+\;k_{1}M_{2}\rangle.
\end{equation}
 The variables in these states can {\it not} be considered as referring to two degrees of
  freedom. Indeed they  are well defined whether or not $M_1$ and $M_2$ are relative
  primes  (in the latter case, of course, $N_1,\;N_2$ do not exist).\\

\begin{equation}
|q_1 k_2;E_{1}>\;\equiv \; { 1 \over \sqrt M_{1}}\Sigma_{k_1}\;e^{-i{2 \pi \over
M_{1}}k_{1}q_{1}}|k_{2}\;+\;k_{1}M_{2}>.
\end{equation}
 It is easily shown that
$$e^{i{ 2\pi \over M_{1}}x}|q_1 k_2;E_{1}>\;=\;e^{i{ 2\pi \over
M_{1}}q_{1}}|q_1 k_2;E_{1}>;\;\;e^{ipL_{2}}|q_1 k_2;E_{1}>\;=\;e^{i{ 2\pi \over
M_{2}}k_{2}}|q_1 k_2;E_{1}>.$$
 Thus this wave function can differ at most by phase from
$|q_1 k_2;C_{1}>$ . This phase difference is gotten by  evaluation of the overlap. In
the following we suppress the (common) eigenvalues indices e.g. $|q_1
k_2;C_{1}>\;\rightarrow\;C_{1}>$ :

$$<C'_{1}|E_{1}>\;=\;{1 \over
M_{1}}\sum_{k'_{1},k_{1}}<k'_{1}N_{1}L_{1}\;+\;k'_{2}N_{2}L_{2}|k_{2}\;+\;k_{1}L_{1}>\exp{[i{2
\pi \over M_{1}}(k'_{1}q'_{1}N_{1}\;-\;k_{1}q_{1})]},$$
$$=\;\Delta^{M_{2}}(k_{2}-k'_{2})\Delta^{M_{1}}(q_{1}-q'_{1})e^{i{2
                \pi \over M_{1}}k_{2}q_{1}N_{1}},$$

The phase difference between $|C_{1}>$ and $|C_{2}>$ may be obtained by evaluating
the overlap between these two wave functions,
$$<C'_{1}|C_{2}>\;=\;{1 \over \sqrt M}\sum_{k'_{1},q_{2}}e^{i{2 \pi \over
M_{1}}k'_{1}q'_{1}N_{1}}<k'_{1}N_{1}L_{1}...|q_{1}N_{1}L_{1}....>e^{i{2 \pi \over
M{2}}k_{2}q_{2}N_{2}}.$$
 Inserting
$$<k'_{1}N_{1}L_{1}...|q_{1}N_{1}L_{1}....>\;=\;{1 \over \sqrt {M}}\exp{-i[{2 \pi \over
M_{1}}(k'_{1}q_{1}N_{1})\;+\;{2 \pi \over M_{2}}(k'_{2}q_{2}N_{2})]}$$

We get
$$<C'_{1}|C_{2}>\;=\;\Delta^{M_{1}}(q_{1}-q'_{1})\Delta^{M_{2}}(k_{2}-k'_{2}).$$
Thus confirming Eq. (\ref {phased}). This can be checked by evaluating the overlap
$<C'_{2}|E_{1}>$: it gives the same result that we got above
for $<C'_{1}|E_{1}>.$\\
 We now consider
 \begin{equation}
|E_{2}>\;\equiv \; {1 \over \sqrt M_{2}}\sum_{q_{2}}e^{i{2 \pi \over
M_{2}}k_{2}q_{2}}|q_{1}\;+\;q_{2}L_{2}>.
\end{equation}
One can check directly that
$$e^{i{2 \pi \over M_{1}}x}|E_{2}>\;=\;e^{i{2 \pi \over M_{1}}q_{1}}|E_{2}>;
\;\;e^{ipL_{2}}|E_{2}>\;=\;e^{i{2 \pi \over M_{2}}k_{2}}|E_{2}>.$$

Thus this function may differ from those above by phase only. We calculate  the
overlap to get,
$$<C'_{2}|E_{2}>\;=\;\Delta^{M_{2}}(k_{2}-k'_{2})\Delta^{M_{1}}(q_{1}-q'_{1})e^{-i{2
\pi \over M_{2}}k_{2}q_{1}N_{2}}.$$

This implies that the phase difference between $|E_{1}>$ and  $|E_{2}>$ is,
$$e^{-i{2\pi \over M_2}k_{2}q_{1}N_{2}}e^{-i{2\pi \over M_{1}}k_{2}q_{1}N_{1}}.$$
We now evaluate this overlap directly to confirm this result,
\begin{equation}\label{E}
<E'_{1}|E_{2}>\;=\;{1 \over \sqrt M}\sum_{k'_{1},q_{2}}e^{i{2 \pi \over
M_{1}}k'_{1}q'_{1}}<k'_{2}\;+\;k_{1}L_{1}|q_{1}\;+\;q_{2}L_{2}>e^{i{2 \pi \over
M_{2}}k_{2}q_{2}}.
\end{equation}
 The overlap,

$$<k'_{2}\;+\;k'_{1}L_{1}|q_{1}\;+\;q_{2}L_{2}>\;=\;{1 \over \sqrt M}e^{-i{2 \pi \over
M}(k'_{2}q_{1}+k'_{2}q_{2}L_{2}+k'_{1}q_{1}L_{1})}.$$

Substituting this into the Eq.(\ref{E}) we get
$$<E'_{1}|E_{2}>\;=\;{1 \over M}e^{i{2 \pi \over M}k'_{2}q_{1}}\sum_{k'_{1}}e^{i{2 \pi
\over M_{1}}k'_{1}(q_{1}-q'_{1})}\sum_{q_{2}}e^{i{2 \pi \over
M_{1}}k'_{1}(q_{1}-q'_{1})}\;=\;\Delta^{M_{1}}(q_{1}-q'_{1})\Delta^{M_{2}}(k_{2}-k'_{2})e^{i{2
\pi \over M}k'_{2}q_{1}}.$$ Thereby confirming the previous anticipation.\\



\end{document}